\documentclass[%
 reprint,
 amsmath,amssymb,
 aps,
]{revtex4-2}
\usepackage{hyperref}
\usepackage{multirow}
\usepackage{graphicx}
\usepackage{dcolumn}
\usepackage{bm}
\usepackage{comment}
\usepackage{xcolor}
\usepackage{lineno}

\newcommand{\dzcomment}[1]{\textcolor{black}{#1}}
\begin{document}

\preprint{APS/123-QED}

\title{Axion Dark Matter eXperiment around 3.3~$\mu$eV with Dine-Fischler-Srednicki-Zhitnitsky Discovery Ability }

\author{C. Goodman}
\author{M. Guzzetti}
\author{C. Hanretty}
 \author{L. J Rosenberg}
   \author{G. Rybka}
\author{J. Sinnis}
  \author{D. Zhang}
  \altaffiliation{Correspondence to dzhang95@uw.edu}
  \affiliation{University of Washington, Seattle, Washington 98195, USA}
   
\author{John Clarke}
\author{I. Siddiqi}
  \affiliation{University of California, Berkeley, California 94720, USA}
\author{A. S. Chou} 
 \author{M. Hollister} 
    \author{S. Knirck}
\author{A. Sonnenschein} 
  \affiliation{Fermi National Accelerator Laboratory, Batavia, Illinois 60510, USA}

\author{T. J. Caligiure}
\author{J. R. Gleason}
\author{A. T. Hipp}
\author{P. Sikivie}
\author{M. E. Solano}
\author{N. S. Sullivan}
\author{D. B. Tanner}
\affiliation{University of Florida, Gainesville, Florida 32611, USA}

\author{R. Khatiwada}
\affiliation{Illinois Institute of Technology, Chicago, Illinois 60616, USA}
\affiliation{Fermi National Accelerator Laboratory, Batavia, Illinois 60510, USA}

\author{G. Carosi}
\author{C. Cisneros}
\author{N. Du}
\author{N. Robertson}
\author{N. Woollett}
\affiliation{Lawrence Livermore National Laboratory, Livermore, California 94550, USA}

\author{L. D. Duffy}
  \affiliation{Los Alamos National Laboratory, Los Alamos, New Mexico 87545, USA}

\author{C. Boutan}
\author{T. Braine}
\author{E. Lentz}
\author{N. S. Oblath}
\author{M. S. Taubman}
  \affiliation{Pacific Northwest National Laboratory, Richland, Washington 99354, USA}

\author{E. J. Daw}
\author{C. Mostyn}
  \author{M. G. Perry}
  \affiliation{University of Sheffield, Sheffield S10 2TN, UK}

\author{C. Bartram}
\affiliation{SLAC National Accelerator Laboratory, 2575 Sand Hill Road, Menlo Park, California 94025, USA}

\author{T. A. Dyson}
\author{S. Ruppert}
\author{M. O. Withers}
\affiliation{Stanford University, Stanford, CA 94305, USA}

\author{C. L. Kuo}
\affiliation{SLAC National Accelerator Laboratory, 2575 Sand Hill Road, Menlo Park, California 94025, USA}
\affiliation{Stanford University, Stanford, CA 94305, USA}

\author{B. T. McAllister}
\affiliation{Swinburne University of Technology, John St, Hawthorn VIC 3122, Australia}

\author{J. H. Buckley}
\author{C. Gaikwad}
\author{J. Hoffman}
\author{K. Murch}
  \affiliation{Washington University, St. Louis, Missouri 63130, USA}
  \author{M. Goryachev}
  \author{E. Hartman}
\author{A. Quiskamp}
\author{M. E. Tobar}
\affiliation{University of Western Australia, Perth, Western Australia 6009, Australia}

\collaboration{ADMX Collaboration}

\date{\today}

\begin{abstract}
We report the results of a QCD axion dark matter search with discovery ability for Dine–Fischler–Srednicki–Zhitnitsky~(DFSZ) axions using an axion haloscope. Sub-Kelvin noise temperatures are reached with an ultra low-noise Josephson parametric amplifier cooled by a dilution refrigerator. This work excludes (with a 90\% confidence level) DFSZ axions with masses between 3.27 to 3.34~$\mu$eV, assuming a standard halo model with a local energy density of 0.45~GeV/cm$^3$ made up 100\% of axions. 
\end{abstract}

\maketitle



Numerous cosmological and astrophysical observations provide compelling evidence for the existence of dark matter, contributing about 85\% of the total mass in our universe~\cite{zwichy1933,vera1970,planck}. Identification of dark matter is one of the outstanding problems in modern particle and astrophysics. 
{\color{black}While a number of particle candidates have been proposed for dark matter such as weakly interacting massive particles~\cite{lz2023,xenon2023,pandax2022}, fuzzy dark matter~\cite{ultralight,fuzzyDM} and sterile neutrinos~\cite{dw,numsm,nustarM31}, the QCD axion~\cite{strongCP,strongCP4,strongCP5}  is particularly well motivated.}


The QCD axion is a consequence of a beyond-standard-model $U(1)$ symmetry introduced to solve the strong CP problem, which, additionally, could comprise some or all of the dark matter~\cite{strongCP,strongCP4,strongCP5}. Measurements of the neutron electric dipole moment place an upper limit on the strong interaction CP violation phase $\theta < 5\times10^{-11}$~\cite{psi2020,axionReview2022}. A global axial $U(1)_{\rm PQ}$ symmetry 
introduced by Peccei and Quinn~\cite{strongCP} undergoes spontaneous symmetry breaking at a very high temperature $T_{\rm PQ}$~\cite{axionReview2022}. At temperatures lower than $T_{\rm PQ}$, a pseudo Nambu-Goldstone boson, the so-called QCD axion, is produced. The mass of the QCD axion will be stabilized after the temperature cools to the QCD phase transition temperature ($\sim$200~MeV). If $T_{\rm PQ}$ is lower than the reheating temperature  after the universe undergoes inflation, QCD axion mass between $\mathcal{O}(1~{\rm \mu eV})$ and $\mathcal{O}(1~{\rm m eV})$ is strongly motivated~\cite{axionReview2022}. 

The Axion Dark Matter eXperiment (ADMX) has been described in previous publications~\cite{admx2001,admx2004,admx2010,admx2016,admx2018,admx2020,admx2021}. The axion haloscope was first proposed by Pierre Sikivie~\cite{pierre} to search for axion dark matter 
decay via the inverse Primakoff effect~\cite{primakoff}, stimulated by immersing the cavity in a strong magnetic field. The outgoing photon carries the rest-mass and kinetic energies of the axion. When the putative photons are on resonance with the cavity modes, the signal is enhanced by the quality factor of the resonator. The enhanced signal is extracted with an antenna and recorded through an ultra-low noise radio frequency~(RF) readout chain.  Previous works have described ADMX's sensitivity to both axion coupling benchmark models: Kim-Shifman-Vainshtein-Zakharov (KSVZ)~\cite{ksvz1,ksvz2,admx2001,admx2004,admx2010,admx2016} and Dine-Fischler-Srednicki-Zhitnitskii (DFSZ)~\cite{dfsz1,dfsz2,admx2018,admx2020,admx2021} over a range of potential axion masses.


In this Letter, we report the exclusion of the axion-photon coupling under the assumption that axions make up all of the dark matter from data collected by ADMX in 2022.  We also include a discussion on the difference between reported exclusion bounds and discovery potential for haloscopes.

{\color{black}The 2022 run} scanned from 792 to 807~MHz (3.27 to 3.34~$\rm \mu eV$ axion mass) to extend the sensitivity reported in the previous run~\cite{admx2021} to DFSZ coupled axions at nominal dark matter densities. {\color{black}The total science data taking spans 71 days with 60\% live time.}


The experimental hardware described in \cite{admx2021} was  refitted with a number of cryogenic upgrades, including replacement of the stainless-steel supports between the 1~K temperature stage and the milliKelvin temperature stage with carbon fiber rods, improved thermal sinking of components,  and additional vacuum space absorption pumping.  These upgrades yielded reductions in
the physical temperatures of the components given in Tab.~\ref{tab:temp}.

\begin{table}
    \caption{
    Cold space temperatures during the previous run and this run. The `attenuator' is the one closest to the cavity that contributes the most thermal noise off-resonance (Fig.~\ref{fig:hfetYfactor}).}\label{tab:temp}
    \begin{ruledtabular}
    \centering
    \begin{tabular}{c c c }
   Location  & {\color{black}2021 run} [mK] & {\color{black}2022 run} [mK] \\\hline
  Attenuator  & 145 & 95 \\
   Cavity & 300 & 130\\
   JPA & 150 & 135 \\
    \end{tabular}
\end{ruledtabular}
\end{table}

\begin{figure}[!htb]
    \includegraphics[width=3.4in]{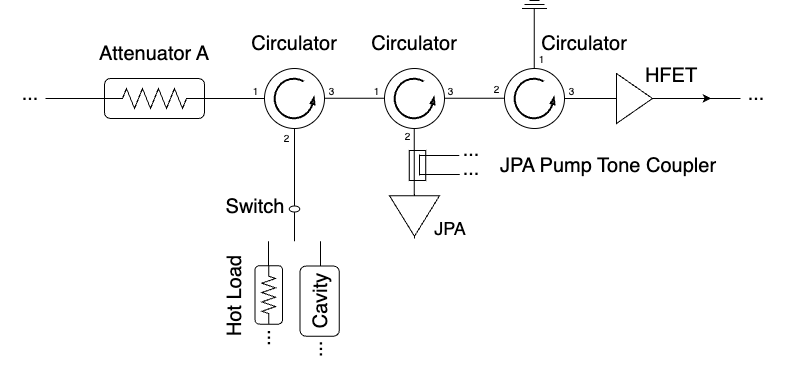}
    \caption{ A schematic of the RF components in the cold space. Other than HFET thermalized to the 4K plate, the components are at the milliKelvin temperature stage.}
    \label{fig:hfetYfactor}
\end{figure}

{\color{black}The signal power of the axion conversion  extracted from the resonant cavity is
\begin{eqnarray}  P_{a}&=&2.8{\times}10^{-23}\mathrm{W} \left(\frac{g_{\gamma}}{0.36}\right)^2 \left(\frac{\rho}{0.45~\mathrm{GeV/ cm^{3}}}\right) 
\nonumber\\ 
& &\left(\frac{f}{800~ \mathrm{MHz}}\right) 
\left(\frac{B}{7.6~ \mathrm{T}}\right)^2 
\left(\frac{V}{106~\mathrm{\ell}}\right)
\left(\frac{C} {0.4}\right)
 \nonumber \\
& &
\left(\frac{Q_{\text{L}}}{45{,}000}\right)\left(\frac{\beta}{1+\beta}\right)\mathcal{L}(f, f_0,Q_L).
\end{eqnarray}
Here $g_{\gamma}=0.36 ~(-0.97)$ is a dimensionless coupling constant for the KSVZ~(DFSZ) axion. The axion-to-two-photon coupling $g_{a\gamma\gamma}$ is  $\alpha g_{\gamma}/\pi f_{a}$,  where $\alpha$ is the fine structure constant, and $f_a$ is the PQ symmetry breaking scale. The local dark matter density is noted as $\rho$. The photon frequency $f$ corresponds to the total energy of the axion $f\approx m_a/h$ where $m_a$ is the axion mass and $h$ is the Planck constant. $B$ is the average magnetic field in the cavity and $V=106~\ell$) is the volume excluding the two rods in the cavity. The form factor $C$ describes the the field overlap between the fundamental mode used for the axion dark matter search and the external magnetic field. $Q_{L}$ and $\beta$ are the loaded quality factor and coupling constant of the cavity, respectively.
$\mathcal{L}(f,f_0,Q_L)=1/(1+4(Q_L(f-f_{0})/f)^2)$ 
is the Lorentzian profile of a cavity with resonant frequency $f_0$. }

The experiment sensitivity to axion-like signals is characterized by the signal-to-noise-ratio,
\begin{equation}\label{eq:snr}
    {\rm SNR}=\frac{F P_{a}}{k_{B}T_{\rm sys}}\cdot \sqrt{\frac{t}{b}},
\end{equation}
where $F$ is the signal efficiency, $T_{\rm sys}$ is the system noise temperature, $k_{B}$ is the Boltzmann constant, $t$ is the integration time, $b$ is the bandwidth of the measured noise power, 
{\color{black}Typical values are  $t=100$~s, $b=100$~Hz for ADMX Run 1~\cite{run1bdetail}.}

As Eq.~\ref{eq:snr} demonstrates, sensitivity depends strongly on $T_{\rm sys}$, which is a combination of the thermal noise from the physical temperature of the cavity and RF components combined with electronic noise introduced by the amplification stages. 
{The overall system noise temperature is $\bar{T}_{sys}=0.48\pm 0.05~{\rm K}$ for the data taken between 792 and 807~$\rm MHz$. \color{black}About half of $T_{\rm sys}$ is the added noise from the first stage Josephson Parametric Amplifier (JPA). The details of the noise calibration are in the supplemental material A~\cite{supplementalLink} (see also references \cite{run1bdetail,NoiseCal,jpaStandingWave} therein).}

 The JPA is a nonlinear device that is biased to operate in a nearly linear manner over a limited range of input pump power levels.  At too high a power, its gain will be compressed and the SNRI mismeasured too low.  For some excitation powers its gain can be mismeasured to be too high due to nonlinear effects.  JPA gain {\color{black}$G_{\rm JPA}$} was measured through both a cavity reflection and transmission measurement which differed in excitation power by about 10 dB.  To ensure that we were operating in the linear regime, only data where the measured gains were consistent between reflection and transmission measurements were used for science analysis. {\color{black} The survived data presents $\bar{G}_{\rm JPA}=21\pm1~{\rm dB}$.}


%


We follow the data taking procedure and analysis established in~\cite{admx2021}, where 100-second power spectra in the vicinity of the ${\rm TM}_{010}$ mode are taken at each tuning rod position and combined to search for candidate axion dark matter signals. {\color{black}
Due to the degraded cavity tuning system, we also developed specific data quality cuts for this run based on the uncertainties of $Q_{L}$ and $\beta$ which replace those used in \cite{admx2021} (see the supplemental material B~\cite{supplementalLink} and references \cite{admx2016, HotzThesis, uncertainty1.5, admx_sidecar_2021, comsol} therein). The fractional systematic uncertainties in SNR are summarized in  Tab.~\ref{tab:uncer}.}


\begin{table}
    \caption{The fractional systematic uncertainties in the $\rm SNR$. \dzcomment{We updated the uncertainty budget for $B^2 VC$ from 3\%~\cite{admx2021} to 6~\% to account for the possible misalignment between the rod and the cavity wall (see the supplemental material B~\cite{supplementalLink}).} Because each digitization has different uncertainties in $Q_{L}$, $\beta$ and $T_{\rm sys}$, we report the median fractional uncertainty.}\label{tab:uncer}
    \begin{ruledtabular}
    \centering
    \begin{tabular}{c c}
    Name  & Fractional uncertainties \\\hline
    $B^2 VC$ & 6\%\\
   $Q_{L}$ (median)&  2\%\\
   $\beta/(1+\beta)$ (median) & 1\%\\
  $T_{\rm sys}$ (median) & 10\% \\
    \end{tabular}
\end{ruledtabular}
\end{table}

The cold receiver transfer function (frequency-dependent gain) in each power spectrum is removed by a sixth-order Pad\'e-approximant first, followed by a second round removal using the averaged residual receiver shapes. In detail, we group every 1000 Pad\'e-fit filtered spectra by time stamps. Then, the averaged residual receiver shape includes 15\% to 85\% quantiles of the 1000 mean ranked in the same group, excluding spectra with transient interference and potential signals. 
The second receiver shape removal avoids mis-identifying the pile-up of the Pad\'e fit residues as excesses during spectrum combination.


{\color{black} The removal of the receiver transfer function reduces the SNR of axion candidates by about 20\% (corresponding to $F=0.8$ in Eq.~\ref{eq:snr}). We quantify the impact of this background removal procedure and subsequent analysis steps by injecting software synthetic signals into the unprocessed power spectra and measuring the detection efficiencies. This Monte Carlo process also confirms our stated confidence level (CL). The software synthetic signals are calibrated via an understanding of our system noise and cavity  parameters including $Q_{L}$ and $\beta$  which are all monitored throughout data-taking with periodic measurements.}

We evaluate the resultant flattened spectra for two signal distributions: a boosted Maxwell-Boltzmann (MB) distribution with dark matter density $0.45~{\rm{GeV/cm^3}}$~\cite{turner,localDMdensity,deSalas_2021}, and the N-body simulated model presented in \cite{lentz} with dark matter density $0.63~{\rm{GeV/cm^3}}$.  
We apply the {\color{black}frequency-dependent} signal shape, accounting for the Lorentzian cavity resonant enhancement, via an optimal filter on each power spectrum, and {\color{black}add the signal strengths weighted by uncertainties } to form one grand spectrum with 100-Hz resolution~\cite{run1bdetail}. 
The systematic uncertainties in the individual frequency bin powers are propagated according to the weights ~\cite{EDthesis}. 

\begin{figure}[!htb]
    \includegraphics[width=3.4in]{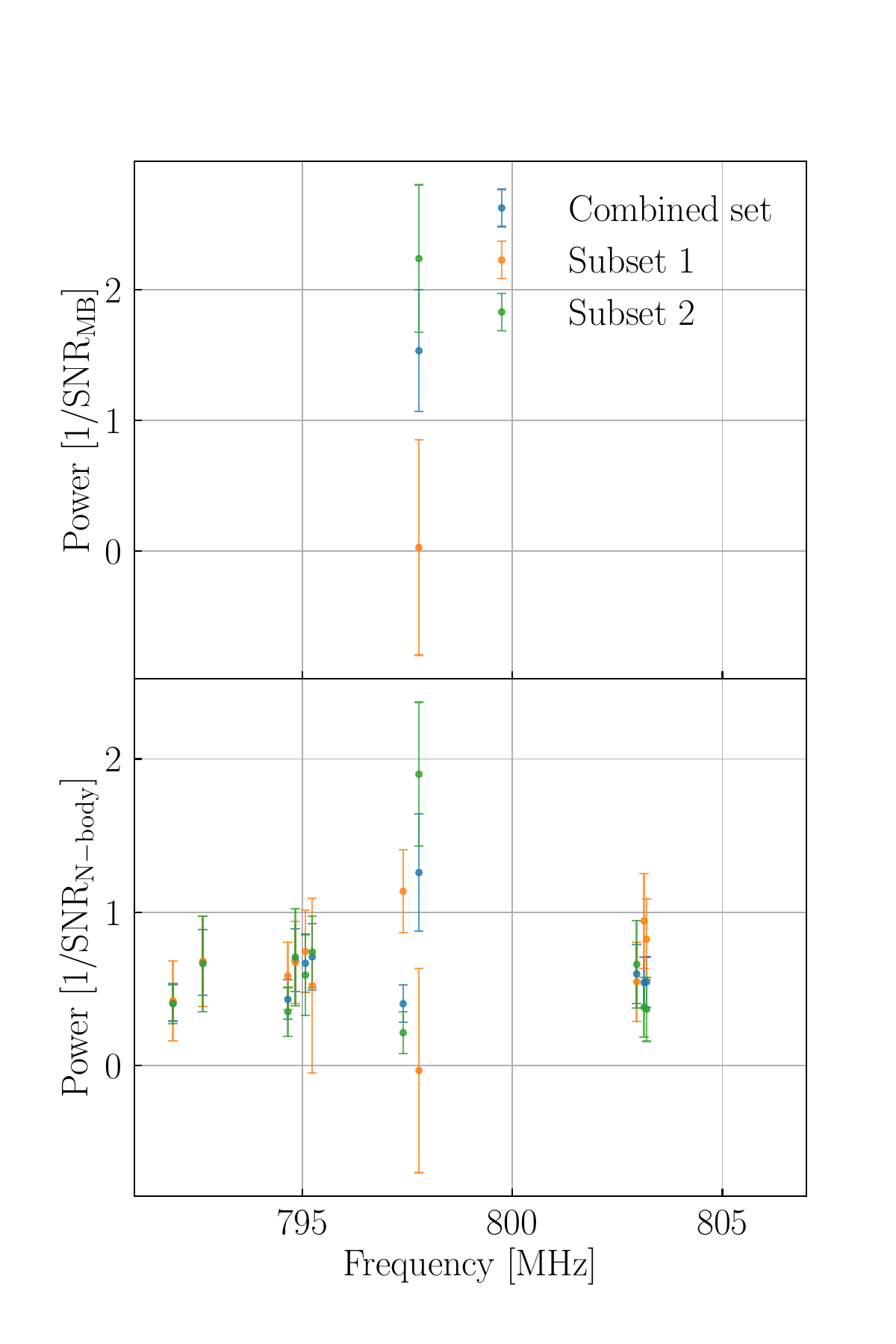}
    \caption{SNR Normalized power of MB (top panel) and N-body (bottom panel) candidates with 3$\sigma$ combined excess (blue). The two subsets (orange, green) power are overlain. The subset data with smaller SNR are preserved for impersistent candidates. The N-body case presents more candidates because the line-width of N-body axion dark matter is narrower, leading to a larger number of independent N-body axion dark matter candidates being scanned and more candidates statistically present as 3$\sigma$ upper fluctuations.
    }
    \label{fig:persist}
\end{figure}

The resultant grand spectrum is searched for excesses above the noise (with definitions discussed below) that would indicate candidate axion signals.  Our initial scan acquired sufficient data such that a DFSZ axion of the MB shape would have an expected ${\rm SNR}\geq3.5$.  Bins that have either a 3$\sigma$ excess or have power measurements greater than $P_{\rm DFSZ}-1.281\sigma$ where $\sigma=1/{\rm SNR}$ when $P_{\rm DFSZ}$ is normalized to 1 (see Fig.~\ref{fig:persist}) are flagged as candidate signals~\cite{run1bdetail}.  At the nominal SNR, the latter criteria is more stringent, so that 90\% of DFSZ-power axion-like signals will be flagged as axion candidates, but the former can be strong in cases where the SNR is larger than our minimum SNR requirement.  
Statistically $0.1\%$ of noise bins are expected to be flagged as candidates with ${\rm SNR}=3.5$.

\begin{figure}[!htb]
    \includegraphics[width=3.4in]{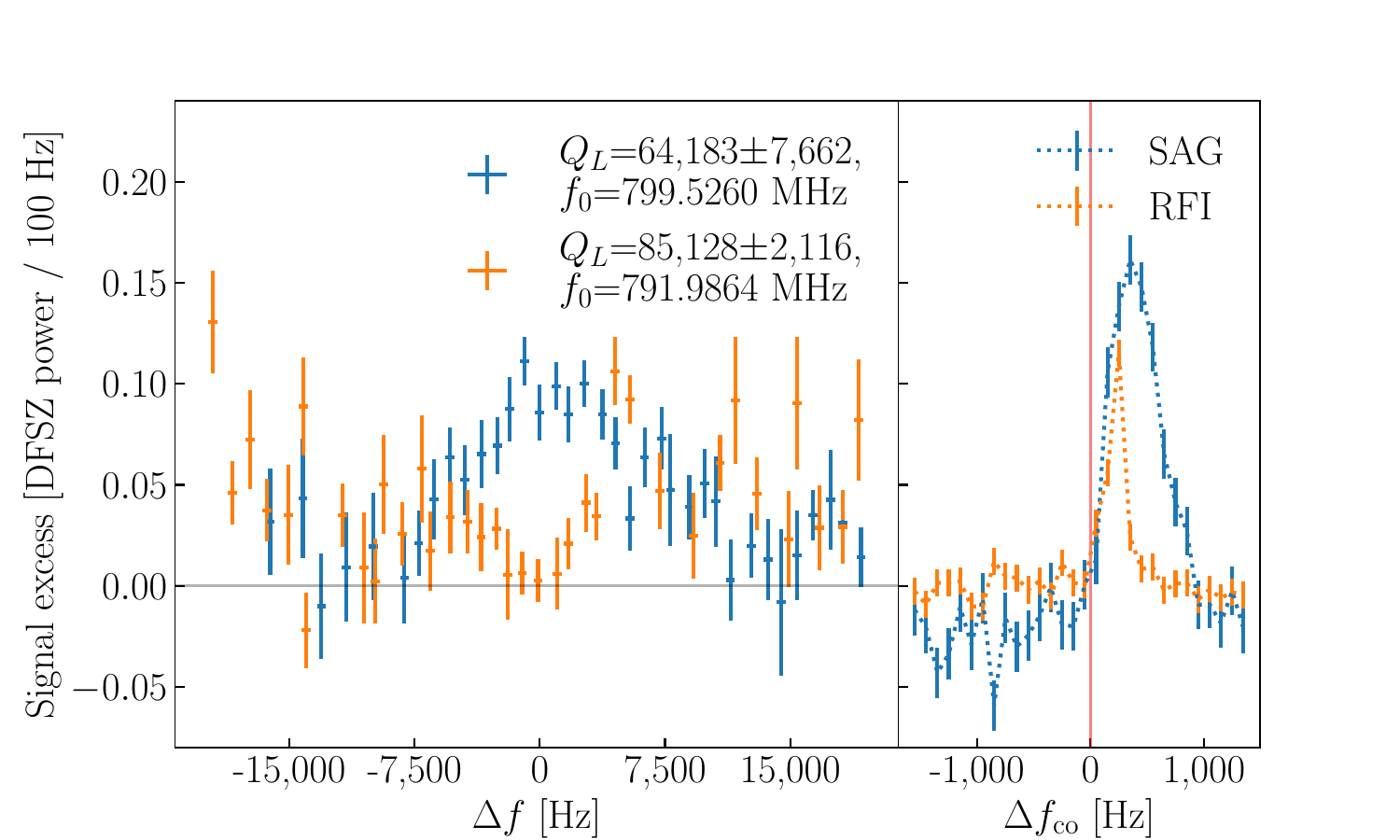}
    \caption{ A SAG signal (blue) found in the initial scan and a persistent RFI (orange) where $f_0$ is the candidate frequency and $\Delta f$ shows the deviation from $f_0$. Left: power excess per 100 Hz (receiver shape divided out) which takes the average of the power excess starting from the identified candidate frequency to 1~kHz beyond. We take the weighted average for the power spectra located in the same frequency bin~\cite{run1bdetail}. 
    Right: weight-summed power excess which is a zoomed-in version of the grand spectrum discussed in the text. Here, $\Delta{f_{\rm co}}=0$ refers to the reference frequencies $f_0$s (SAG only includes the initial scan, and RFI includes both the initial and re-scans). How the excesses change from off resonance to on resonance of the cavity is not included on the weight-summed power excesses (right) where the SAG and RFI are difficult to separate. However, if we group the power spectra according to the deviation to the cavity resonance, the grouped averaged power excesses (left) of the SAG are enhanced on resonance of the cavity and on the contrary, those of the RFI are reduced.  
    }
    \label{fig:sagAndRfi}
\end{figure}

\begin{figure}[!htb]
    \includegraphics[width=3.5in]{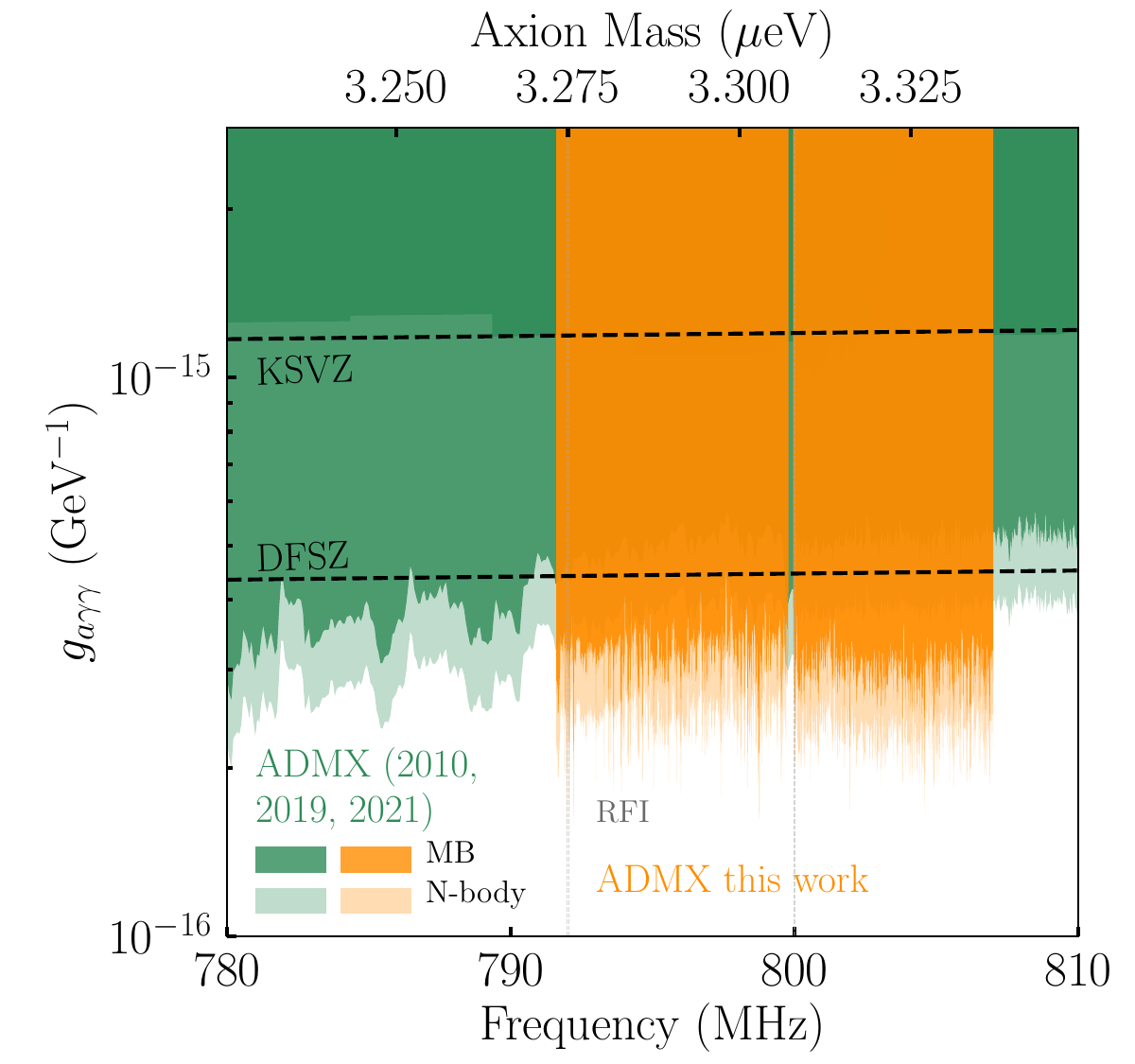}
    \caption{The exclusion limits on $g_{a\gamma\gamma}$ with 90\% CL of this work (orange shadow for the MB distribution and semi-transparent shadow for the N-body distribution). The previous ADMX searches with the same axion masses are overlain with dark green shadows~\cite{admx2010}, light green shadows~\cite{admx2020,admx2021} for the MB distributions, and the semi-transparent green shadows~\cite{admx2020,admx2021} for the N-body distributions. The RFIs are  marked by the gray dashed lines. The two benchmark $g_{a\gamma\gamma}$ couplings (KSVZ and DFSZ) are marked with the black lines. }
    \label{fig:limit}
\end{figure}

Proper flagging of candidate signals is tested with the injection of axion signal-shaped RF power into the cavity {\color{black}from a weakly coupled port} using a system called the synthetic axion generator~\cite{sag}.  The analysis personnel are blinded to the operation of the SAG during the initial round of data taking and the first re-scan.  Synthetic signals are unblinded after the first re-scan, and the region is re-scanned with the signal turned off, and the signal is verifed as having  disappeared.  All the synthetic signals were identified successfully and excluded with re-scans when the SAG was turned off during our data taking. Other than the SAG signals, we identified three persistent radio-frequency interference~(RFI) signals. A signal coming from the cavity is maximized while on-resonance and follows the line-shape of the resonance, and the width of the signal strength is related to $Q_L$.  Signals picked up by the receiver in components beyond the cavity will either have a strength independent of the cavity resonant frequency, or in some cases maximize only off-resonance of the cavity. 
Representative SAG and RFI signals are shown in Fig.~\ref{fig:sagAndRfi}, with features important to discrimination indicated. All the three persistent RFI signals are marked in Fig.~\ref{fig:limit} with gray dashed lines.


\begin{table}[htb]
    \caption{
    The number of candidates (with 100~Hz as the frequency resolution) not reaching to DFSZ 90\% CL exclusion or presenting three sigma ($3\sigma$) excesses in this run and with previous data sets combined. All the $3\sigma$ excesses for both scenarios fail the persistence check as described below.}\label{tab:combine}
    \begin{ruledtabular}
    \centering
    \begin{tabular}{c c c }
   Distribution  & This & Combined  \\\hline
    MB (not reaching DFSZ) & 3543 & 1181 \\
   N-body (not reaching DFSZ) & 579 & 39 \\
   MB ($3\sigma$) & 4 & 1 \\
   N-body ($3\sigma$) & 19 & 11\\
    \end{tabular}
\end{ruledtabular}
\end{table}


Candidate signals are re-scanned---more power spectra are acquired within $\pm 3.5~$kHz of the candidate frequency, and the persistence of the candidate is evaluated with the procedure defined in \cite{run1bdetail}. Typically enough integration time was spent on each candidate for a DFSZ signal to have a 4$\sigma$ significance, thus clearly identifying signals and rejecting noise.  We distinguish here between our upper bound on $g_{a\gamma\gamma}$ depicted in Fig.~\ref{fig:limit} as the exclusion limit, and the likelihood we would have successfully flagged and re-scanned a signal DFSZ axion at the nominal dark matter density (our discovery potential).


Re-scanned candidates that did not persist can be trivially eliminated as described in Ref.~\cite{admx2021}.  However, as a mechanical fault ended the data taking in the midst of the re-scan process, and some calibrations and adjustments to analysis were done after the end of the run, there remained some regions of either insufficient sensitivity or un-rescanned 3$\sigma$ excesses that needed to be addressed.

Data from this run were combined with the data from \cite{admx2020,admx2021} for computation of our overall sensitivity.  Any regions still with insufficient sensitivity to exclude a DFSZ axion signal at 90\% CL were treated as gaps in the data, and the local exclusion bounds were de-weighted accordingly.
Frequency regions that still had a 3$\sigma$ excess
were examined for persistence by dividing the related spectra into two subsets with respect to timestamps as illustrated in Fig.~\ref{fig:persist}. 
In all cases the two subsets were inconsistent with a persistent axion signal. These therefore can either be a transient or a statistical fluctuation. 
We used only the conservative subset with the lower SNR to set bounds on the axion-photon coupling, and do not exclude the possibility the excesses can be more prominent with higher SNRs.
The effects of data combination are summarized in Tab.~\ref{tab:combine}.

As we had no persistent axion signal candidates, we interpreted the combined measured excess power in every bin as potentially due to an axion signal and computed the Frequentist upper limit at 90\% CL as defined in Ref.~\cite{10.1093/ptep/ptaa104} for each bin {\color{black}(see the Appendix A for a wider mass range)}.  These limits are smoothed for visualization in Fig.~\ref{fig:limit}.  
Most bins exclude $g_{a\gamma\gamma}$ extensively below the level predicted for a DFSZ axion because we target ${\rm SNR}_{\rm MB}=3.5$ in the initial data taking to have the ability to further study the gaps and potential signals. Since lower couplings would not have necessarily been flagged for re-scan, we claim a discovery potential of only $90\%$ for DFSZ axion signals.

In summary, 
we set the most sensitive exclusion limits of $g_{a\gamma\gamma}$ with 90\% CL as a function of the axion mass between 3.27 and 3.34~$\rm \mu eV$~(Fig.~\ref{fig:limit}), 
excluding the DFSZ axion as the dominant local dark matter with such masses with the same CL.

\begin{acknowledgments}
We thank Jason Detwiler at University of Washington for discussions on the unbiased uncertainty estimator. Chelsea Bartram acknowledges support from the Panofsky Fellowship at SLAC. This work was supported by the U.S. Department of Energy through Grants No DE-SC0009800, No. DE-SC0009723, No. DE-SC0010296, No. DE-SC0010280, No. DE-SC0011665, No. DE-FG02-97ER41029, No. DE-FG02-96ER40956, No. DE-AC52-07NA27344, No. DE-AC03-76SF00098, No. DE-SC0022148 and No. DE-SC0017987. This manuscript has been authored by Fermi Forward Discovery Group, LLC under Contract No. 89243024CSC000002 with the U.S. Department of Energy, Office of Science, Office of High Energy Physics. Pacific Northwest National Laboratory is a multi-program national laboratory operated for the U.S. DOE by Battelle Memorial Institute under Contract No. DE-AC05-76RL01830. University of Sheffield acknowledges the Quantum Sensors for the Hidden Sector (QSHS) Extended Support under the grant ST/Y004620/1. UWA and Swinburne University participation is funded by the ARC Centres of Excellence for Engineered Quantum Systems, Grant No. CE170100009, and Dark Matter Particle Physics, Grant No. CE200100008. Additional support was provided by the Heising-Simons Foundation and by the Laboratory Directed Research and Development offices of the Lawrence Livermore and Lawrence Alamos National Laboratories. PNNL Release No.: PNNL-SA-204925. LLNL Release No.: LLNL-JRNL-871262. LANL release No.: LA-UR-25-21087.
\end{acknowledgments}

\bibliographystyle{apsrev4-2}
\bibliography{apssamp.bib}
\appendix

\section{Axion-photon Coupling Limit}
\begin{figure*}[!htb]
    \includegraphics[width=6in]{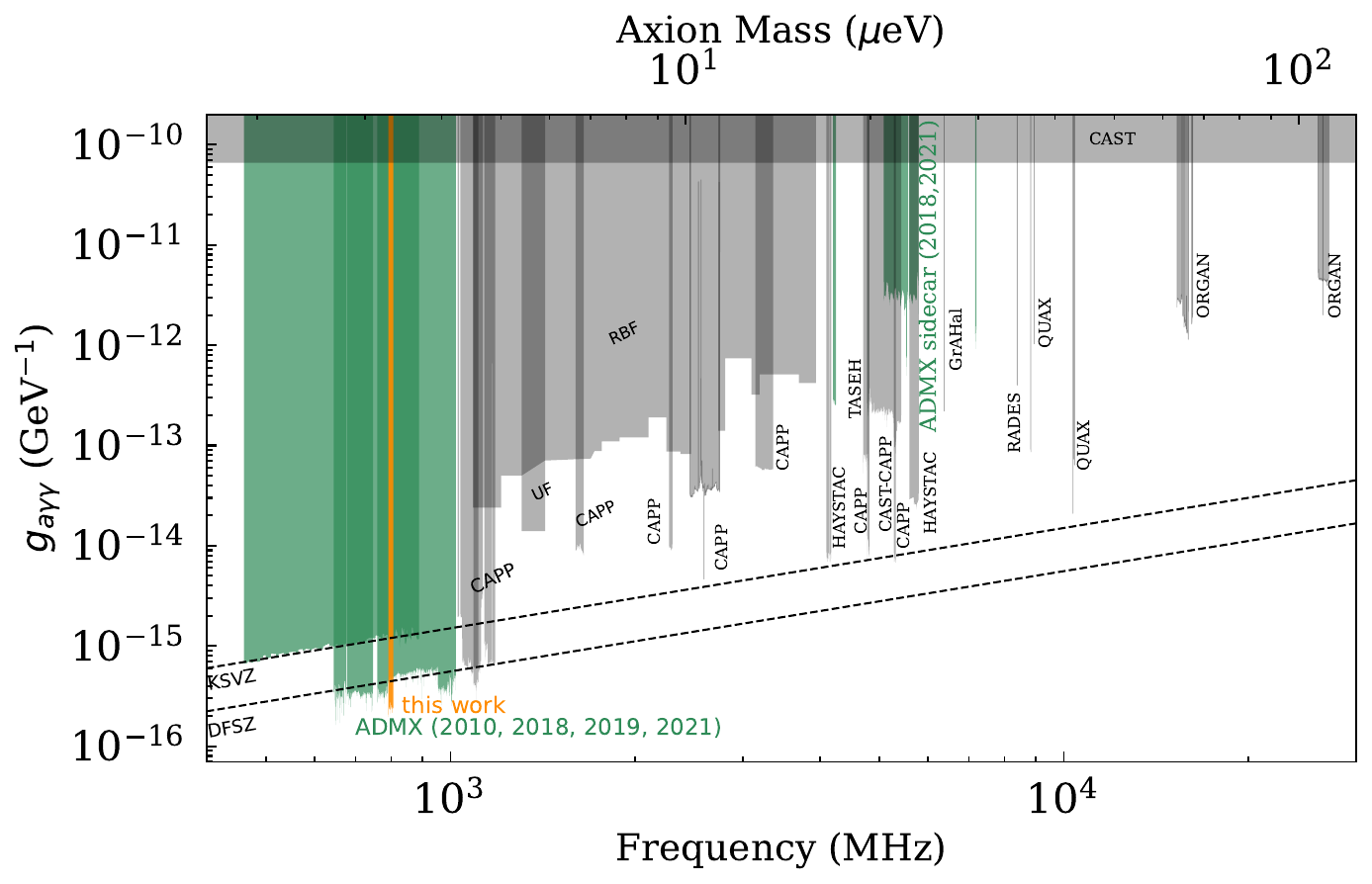}
    \caption{The exclusion limits on $g_{a\gamma\gamma}$ with 90\% CL of this work with the standard halo model assumed. The previous $g_{a\gamma\gamma}$ limits from ADMX in the same or nearby axion mass regions are overlain with dark green shadows \cite{admx2010,admx2018,admx2020,admx2021,admx_sidecar_2021,ADMX:2018ogs}. Other limits of axion-photon coupling in the same or nearby region are overlain in grey shadows, including the work from Center for Axion and Precision Physics (CAPP)~\cite{Lee:2020cfj,Jeong:2020cwz,CAPP:2020utb,Lee:2022mnc,Kim:2022hmg,Yi:2022fmn,Yang:2023yry,Kim:2023vpo,CAPP:2024dtx}, University of Florida 1990 (UF)~\cite{uf_limit_1990}, Rochester-Brookhaven-Florida 1989 (RBF)~\cite{Wuensch:1989sa}, Haloscope At Yale Sensitive To Axion Cold dark matter (HAYSTAC)~\cite{HAYSTAC:2018rwy,HAYSTAC:2020kwv,HAYSTAC:2023cam}, Taiwan Axion Search Experiment with Haloscope (TASEH)~\cite{TASEH:2022vvu}, CERN Axion Solar Telescope (CAST)~\cite{cast2024}, CAST-CAPP~\cite{Adair:2022rtw}, Relic Axion Dark Matter Exploratory Setup (RADES)~\cite{CAST:2020rlf}, Grenoble Axion Haloscope project (GrAHal)~\cite{Grenet:2021vbb}, QUaerere AXions experiment (QUAX)~\cite{Alesini:2019ajt,Alesini:2020vny,Alesini:2022lnp,QUAX:2023gop,QUAX:2024fut},  Oscillating Resonant Group AxioN experiment (ORGAN)~\cite{McAllister:2017lkb,Quiskamp:2022pks,Quiskamp:2023ehr}. The two benchmark $g_{a\gamma\gamma}$ couplings (KSVZ and DFSZ) are marked with the black lines. }
    \label{fig:limitWide}
\end{figure*}

Figure \ref{fig:limitWide} shows the exclusion limits on $g_{a\gamma\gamma}$ with 90\% CL of this work and other results in the same or nearby axion mass region.

\end{document}